\documentclass{raa}

\usepackage{graphicx,times}             
\input{epsf.sty}

\newcommand{\lsim}{\raisebox{-.4ex}{$\stackrel{<}{\scriptstyle \sim}$}}
\newcommand{\gsim}{\raisebox{-.4ex}{$\stackrel{>}{\scriptstyle \sim}$}}

\begin{document}

\title{Nucleosynthesis in the accretion disks of Type II collapsars
}

\author{Indrani Banerjee \and Banibrata
Mukhopadhyay} 

\institute{Department of Physics,
Indian Institute of Science, Bangalore 560012, India; 
{\it indrani@physics.iisc.ernet.in, 
bm@physics.iisc.ernet.in} \\ 
}


\abstract{
We investigate nucleosynthesis inside the gamma-ray burst (GRB) accretion disks formed by the Type II 
collapsars. In these collapsars, the core collapse of massive stars first leads to the formation of a 
proto-neutron star. After that, an outward going shock launches a successful supernova. However, this 
supernova ejecta lack momentum and within a few seconds this newly formed neutron star gets transformed to 
a stellar mass black hole via massive fallback. The hydrodynamics of such an accretion disk formed from the 
fallback material of the supernova ejecta has been studied extensively in the past. We use these well 
established hydrodynamic models for our accretion disk in order to understand
nucleosynthesis, which is mainly advection dominated in the outer 
regions. Neutrino cooling becomes important in the inner disk where the temperature and density are higher.
Higher the accretion rate ($\dot{M}$), higher is the density and temperature in the disks.
In this work we deal with accretion disks with relatively low accretion rates: $0.001 M_{\odot} 
s^{-1} \lsim \dot{M} \lsim 0.01 M_{\odot} s^{-1}$  
and hence these disks are predominantly advection dominated. 
We use He-rich and Si-rich abundances as the initial condition of nucleosynthesis at the outer disk, and 
being equipped with 
the disk hydrodynamics and the nuclear network code, we study the abundance evolution as matter inflows and 
falls into the central object. We investigate the variation in the nucleosynthesis products in the disk 
with the change in the initial abundance at the outer disk and also with the change in the mass accretion 
rate. We report the synthesis of several unusual nuclei like $^{31}$P, $^{39}$K, $^{43}$Sc, $^{35}$Cl, 
and various 
isotopes of titanium, vanadium, chromium, manganese and copper.
We also confirm that isotopes of iron, cobalt, 
nickel, argon, calcium, sulphur and silicon get synthesized in the disk, as shown by previous authors. 
Much of these 
heavy elements thus synthesized are ejected from the disk via outflows and hence they
should leave their signature in observed data.
\vskip0.5cm
{\bf
\noindent Key~words:~~accretion, accretion disk --- gamma rays: bursts --- collapsars --- nucleosynthesis --- abundance
}
}

\authorrunning{Banerjee \& Mukhopadhyay}

\titlerunning{Nucleosynthesis in the accretion disks of Type II collapsars}

\maketitle

\section{Introduction}

With the advent of the Italian-Dutch satellite BeppoSAX and the BATSE detectors on the Compton Gamma-Ray 
Observatory, there has been a spectacular increase in the data available for studying long-duration 
gamma-ray bursts (GRBs). The availability of these data has sparked the idea that these events are 
associated 
with core-collapse supernova (SN). Usually these core collapse supernovae (SNe) are categorized under Type 
Ib or Type Ic SNe. Type Ib SNe result when the collapsing star has lost most of its hydrogen envelope. When 
the progenitor star loses most of its helium envelope along with the outer hydrogen envelope, it leads to 
Type Ic SNe. Such a core collapse often leads to the formation of a black hole with a high density, 
differentially rotating accretion disk around it. Rotating conducting fluids generate magnetic fields and 
differential rotation twists the magnetic field lines violently which causes a jet of material to blast 
outward perpendicular to the accretion disk. The watershed event that first established the SN-GRB 
connection was the discovery of GRB 980425 in association with one of the most unusual SNe ever seen, SN 
1998bw (Galama et al. 1998). After that, many more such events followed, namely, the association of SN 
2003dh with GRB 030329 (Hjorth et al. 2003), SN 2003lw with GRB 031203 and SN 2002lt with GRB 021211 (for a 
review see Woosley \& Bloom 2006). 

However, not all SNe of Type Ibc are accompanied by long duration soft spectrum GRBs. 
Therefore, it is necessary to understand when do the core collapse of massive stars lead to the GRBs. It 
has been hypothesized that GRBs result from most massive and most rapidly rotating low metallicity 
stars. This is because only very massive stars can undergo core collapse to form black holes. The rapid 
rotation endows it the requisite angular momentum so that an accretion disk develops around the black hole 
and the low metallicity enables the star to strip off its hydrogen envelope easily so that the jets that 
are formed can pummel through the star, reach the stellar surface and finally break out into the 
surroundings as GRBs (Heger et al. 2003).

The collapsar model is the most promising theoretical model explaining the long-duration GRBs 
and the SN-GRB connections (MacFadyen \& Woosley 1999,
hereafter MW99, MacFadyen et al. 2001). It explores the end states of stellar collapse of very massive 
stars with mass in the main sequence ($M_{MS}$) at least greater than $20M_{\odot}$ where $M_{\odot}$ is the 
mass of the Sun. Two types of collapsar models 
have been proposed. Type I collapsar models the core collapse of 
very massive stars with $M_{MS} \gsim 40 M_{\odot}$. No supernova explosion takes place during such a 
core 
collapse event and hence they are termed as ``failed supernovae'' (Fryer 1999). 
In these collapsars, a black hole forms within a few seconds of the onset of the iron core collapse. 
If the progenitor star has 
sufficient 
angular momentum, the range being predicted by Heger et al. (2000), a quasi-steady accretion 
disk develops around the black hole several seconds after the core collapse (MW99). In these disks very
 high accretion rates with $\dot{M}  \gsim  0.1  M_{\odot} s^{-1}$ is maintained for approximately 10-20 s 
 (MW99). For such high accretion rates, the temperature of the accretion disk rises much above $10^{10}$K 
 so  that the accreting gas contains only protons and neutrons (Popham et al. 1999). Therefore, the 
 progenitors 
 with $M_{MS} \gsim  40 M_{\odot}$ are unimportant for the synthesis of heavy elements in the accretion 
 disk.

Second type of collapsars, namely Type II collapsars or ``fallback collapsars", are formed from the core 
collapse of progenitors with $20 M_{\odot}\lsim M_{MS}<40 M_{\odot}$. A proto neutron star is formed 
immediately after this core collapse. Then an outward going shock launches a supernova successfully and the 
nascent neutron star is soon transformed into a black hole due to massive fallback of the supernova ejecta 
(Fryer 1999). With the increase in the progenitor mass the amount of supernova ejecta falling back onto the 
neutron star increases (Woosley \& Weaver 1995, Fryer 1999, Fryer \& Heger 2000). The matter accreting, in 
these cases, onto the black hole forms a disk with accretion rate 1-2 orders of magnitude less than that of 
Type I collapsars. Since the accretion rate is lower in this case, the density and temperature in these 
disks are also lower and ideal for the nucleosynthesis of heavy elements.

In this paper, we focus our attention on the nucleosynthesis of heavy elements in GRB accretion disks and 
hence consider the accretion disks formed by the fallback collapsars. We report the synthesis of various 
 elements like $^{31}$P, $^{39}$K, $^{43}$Sc, $^{35}$Cl and several  
uncommon isotopes of titanium, vanadium, chromium, manganese and copper in the disk for the first time, 
apart from 
isotopes of iron, 
cobalt, nickel, calcium, argon, sulphur and silicon which have already been reported earlier
(e.g. Matsuba et al. 2004). 

The paper is organized as follows. In the next section we present the hydrodynamical model and the input 
physics we use for the GRB accretion disk. In \S 3 we describe the nuclear network to calculate the change 
of abundance of elements in the disk along with the initial abundances of the elements we begin with at the 
outer disk. We follow the abundance evolution in the disk in \S 4. Finally 
we briefly summarize the results in \S 5.  

\section{Disk Model and Input Physics}

Let us briefly recapitulate the underlying disk physics recalling the results from the existing 
literatures.
We model an accretion disk formed in a Type II collapsar within the framework suggested by previous authors 
(Matsuba et. al 2004, Kohri et. al 2005, hereafter KNP05, Chen \& Beloborodov 2007, hereafter CB07). For 
the present purpose, we ignore the 
time dependent properties of the flow and focus mainly on the overall properties of the disk.
 
Hence, for simplicity we construct a steady, axisymmetric model of the accretion disk around a stellar mass 
black hole with accretion rates upto $0.01 M_{\odot} s^{-1}$ which is the characteristic of accretion disks 
in Type II collapsars. 
\subsection {Neutrino emission and opacities}
As the accretion rate is very high in these disks, the density and temperature of the inflowing gas 
reach their values of the order of
$10^{10} \rm gcm^{-3}$ and $10^{10} \rm K$, respectively, near the inner edge of the disk (e.g. CB07). Many 
interactions which lead to the emission of neutrinos gets turned on in this regime. 
Such disks are often termed as neutrino dominated accretion flows (NDAFs) in the literature. There are four 
different mechanisms of neutrino emission (see, e.g., Popham et al. 1999, Di Matteo et al. 2002, hereafter 
DPN02, Kohri \& Mineshige 2002), namely, electron capture by protons: $p + e^-\rightarrow n + \nu_e$ and 
positron capture by neutrons: $n + e^+ \rightarrow p + \overline{\nu}_e$, electron-positron pair 
annihilation: $e^+ + e^- 
\rightarrow \nu_i + \overline{\nu}_i$, where $i$ represents neutrinos of all the three flavors, 
nucleon-nucleon bremsstrahlung: $n + n \rightarrow n + n + \nu_i + \overline{\nu}_i$, and finally plasmon 
decay: 
$\overline{\gamma} \rightarrow \nu_e  + \overline{\nu}_e$. The last two processes become important when 
electron degeneracy is extremely high and hence are negligible in GRB disks.

There is an absorption process corresponding to each neutrino emission process. Additionally, scattering 
of neutrinos off free nucleons prevents the free escape of neutrinos from the disk. 
Above processes have their associated absorptive and scattering optical depths 
(see, e.g., DPN02).

\subsection {Basic assumptions and hydrodynamic equations}
We adopt height-averaged equations based on a pseudo-Newtonian framework. General relativistic effects due
to the black hole are modelled in terms of the gravitational force of the well established pseudo-Newtonian 
potential (Mukhopadhyay 2002) given by
\begin{equation}
\frac{\lambda ^{2}_K}{R^3}=\frac{GM(R^2-(2GMa/c) \sqrt{2GMR/c^4}+G^2 M^2 a^2/c^4)^2}
{R^2(R^2-2GMR/c^2+(GMa/c)\sqrt{GMR/c^4})^2}=F_R,
\end{equation} 
where $R$ is the radial distance of the black hole from an arbitrary flow orbit, $F_R$ the gravitational 
force of the black hole at $R$, which mimics the major features of general relativistic effects important 
for the present purpose quite well for $R \gsim 2R_g$, where $R_g = 2GM/c^2$ is the Schwarzschild radius, 
$M$ is the mass of the black hole, $G$ the Newtonian gravitation constant, $c$ the speed of light, $a$ 
the dimensionless specific angular momentum of the black hole (Kerr parameter), and $\lambda_K$ the 
Keplerian specific angular momentum of the disk. 
Note that self-gravity of the disk under consideration is negligible as long as 
$\dot{M} < 10 M_{\odot} s^{-1}$ for a stellar mass 
black hole. We adopt the cylindrical polar coordinate system $(R,\phi,z)$ assuming the black hole to be
at the origin. We assume that 
$c_s^2 \approx P/\rho$ and the vertical scale height as $H = c_s/\Omega_K$, where $c_s$ is the sound speed,
$P$ the flow pressure, $\rho$ the flow density and $\Omega_K$ = $\sqrt {F_R/R}$ is the Keplerian angular 
velocity.

The continuity equation is given by
\begin{equation}
\dot{M} = 4\pi R \rho H v_R \approx 6 \pi \nu \rho H,
\end{equation}
where $\dot{M}$ is the mass accretion rate which is constant, \textit{$v_R$} is the radial velocity of the 
accreting gas and $\nu$ is the kinematic viscosity 
coefficient given by
\begin{equation}
\nu = \alpha c_s^{2} / \Omega_K.
\end{equation}
Last part of equation (2) is obtained by ignoring a boundary correction term (see e.g. Frank 
et al. 1992).
The $R$- and $\phi$-components of the momentum balance equation are
\begin{equation}
v_R\frac{dv_R}{dR} + (\Omega ^{2}_K - \Omega ^{2})R + \frac{1}{\rho} \frac{dP}{dR}=0 ,
\end{equation}
\begin{equation}
v_R\frac{d(\Omega R ^2)}{dR}= \frac{1}{\rho R H} \frac{d}{dR} (\frac{\alpha \rho c_s^2 R^3 H}
{\Omega_K}\frac{d \Omega}{dR}),
\end{equation}
where $\Omega$ is the angular velocity of the flow and $\alpha$ is the dimensionless viscosity parameter 
(Shakura \& Sunyaev 1973).

In the equation of state, we include the contributions from radiation pressure, baryon pressure, electron 
degeneracy pressure 
and neutrino pressure (Popham \& Narayan 1995, Popham et al. 1999, Narayan et al. 2001,  KNP05, CB07), 
i.e,
\begin{equation}
P = P_\gamma + P_b + P_e + P_\nu .
\end{equation}

The radiation pressure is given by
\begin{equation}
P_\gamma = \frac{1}{3} \overline{a} T^4 ,
\end{equation}
where $T$ is the temperature of the flow in $K$ and $\overline{a}$ the radiation constant.

The baryon pressure is given by
\begin{equation}
P_b = \frac{\rho•}{m_p•} k T \left(X_f + \frac{1-X_f}{4•}\right),
\end{equation}
where $k$ is the Boltzmann constant, $m_p$ the mass of the proton, $X_f$ the free nucleon mass 
fraction and $1-X_f$
the mass fraction of $\alpha$-particles.
The equation of nuclear statistical equilibrium provides $X_f$ (see Meyer 1994,
CB07, for detailed expression). It should be noted that 
while constructing the disk model, we consider only neutrons, protons and 
$\alpha$-particles. However, after obtaining hydrodynamics of the disk, 
we actually evaluate the chemical composition inside the disk in 
detail using the nuclear reaction network which will be mentioned in \S 3.

Electrons are mildly degenerate and sometimes ultrarelativistic in these accretion disks.
Therefore to calculate the electron pressure we have to use the exact Fermi-Dirac distribution 
(KNP05, CB07).
Since electrons and positrons both contribute to the pressure 

\begin{equation}
P_e = P_{e^{-}} + P_{e^{+}},
\end{equation}
where
\begin{equation}
P_{e^{-}} = \frac{1}{3 \pi^2 \hbar^3 c^3•} \int^{\infty} _0 \frac{p^4•}{\sqrt{p^2 c^2 + m_e^2 
c^4}•}\frac{1}{e^{(\sqrt{p^2 c^2 + m_e^2 c^4}-\mu_e )/kT} + 1}\,dp
\end{equation}
and
\begin{equation}
P_{e^{+}} = \frac{1}{3 \pi^2 \hbar^3 c^3•} \int^{\infty} _0 \frac{p^4•}{\sqrt{p^2 c^2 + m_e^2 
c^4}•}\frac{1}{e^{(\sqrt{p^2 c^2 + m_e^2 c^4}+\mu_e )/kT} + 1}\,dp
\end{equation}
and $\hbar$, $m_e$ and $\mu_e$ are reduced Planck's constant, mass of the electron and electron chemical
potential respectively. It is to be noted that the above expression is valid for both relativistic and
non-relativistic electrons. The procedure for calculating the electron chemical potential is discussed
in detail in KNP05 and CB07.   

Finally, the contribution to the pressure due to the neutrinos is given in the forth term of
equation (6). For the expression of neutrino pressure see equation (17) of KNP05.

The energy generated through viscous dissipation in the disk is partly advected with the inflowing gas and 
the remaining is radiated through photons and neutrinos and lost via photodisintegration.
Thus the equation for energy per unit area per unit time of the disk is given by
\begin{equation}
Q^+_{vis}=Q^-_{adv}+Q^-_{rad}+Q^-_{\nu}+Q^-_{photo}+Q^-_{nuc} ,
\end{equation}  
where
\begin{equation}
Q^+_{vis}=\frac{3 G M \dot{M}}{8 \pi R^3},
\end{equation} 
is the heat generated through viscous dissipation.

We approximate $Q^-_{adv}$ by (see, e.g., Narayan \& Yi 1994, Abramowicz et al. 1995, DPN02) 
\begin{equation}
Q^-_{adv}=\Sigma v_{R} T\frac{ds}{dR}
\end{equation} 
where $s$ denotes the entropy per particle and $\Sigma$ the surface density given by
\begin{equation}
\Sigma = (\rho H)/2.
\end{equation}

The entropy per particle $s$ is further given by 
\begin{equation}
s = (s_{rad} + s_{gas})/ \rho , 
\end{equation}
where $s_{rad}$ is the entropy density of radiation given by
\begin{equation}
s_{rad}=\frac{2}{•3} \overline{a} g_* T^3
\end{equation}
$g_*$ being the number of degrees of freedom of the photons which is equal to 2 and $s_{gas}$ is the 
entropy density of gas given by
\begin{equation}
s_{gas}=\sum_i n_i \left\lbrace \frac{5}{2•} + \ln \left[\frac{g_i}{n_i•}\left(\frac{m_i T}{2 
\pi•}\right)^{3/2}\right]\right\rbrace .
\end{equation}
The subscript $i$ runs over nonrelativistic nucleons and electrons and $g_i$ represents the statistical
degree of freedom of species $i$. Since the entropy of degenerate particles is very small it can be safely
neglected.

\begin{figure}
\centering
\includegraphics[width=0.9\columnwidth]{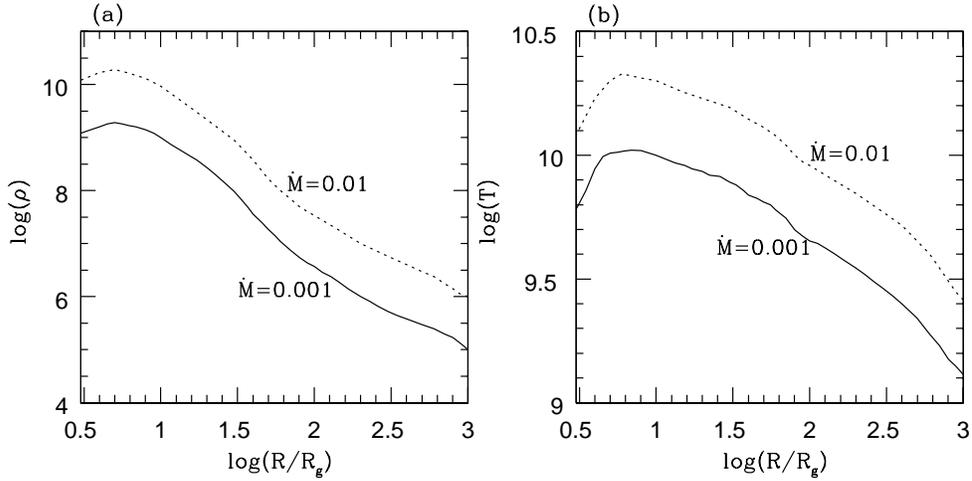}
\caption{
Density and temperature profiles of the accretion disks around a $3 M_{\odot}$ Schwarzschild black 
hole using accretion rate as the parameter, where $r_{g}= 2 G M/c^2$.
 }
\label{figlam} \end{figure}

Then heat radiated per unit area 
\begin{equation}
Q^-_{rad}=\frac{g_* \sigma_s T^4}{2\tau_{tot}},
\end{equation}
where $\sigma _s$ is the Stefan-Boltzmann constant. The optical depth, $\tau_{tot}$, is given by 
\begin{equation}
\tau_{tot}=\kappa_R \rho H = \frac{\kappa_R \Sigma}{2},
\end{equation}  
where the Rosseland mean opacity
\begin{equation}
\kappa_R=0.40 + 0.64 \times 10^{23}\left(\frac{\rho}{g \; cm ^{-3}}\right)\left(\frac{1}
{T/K}\right)^{3}\;g^{-1}\: cm^{2}.
\end{equation}
The first term on the right hand side in equation (21) is from electron scattering and the second term is 
from 
free-free absorption. The temperature and the density profiles in an NDAF are such that the radiative 
optical depth is extremely high and hence the radiative cooling term is negligible compared to the other 
cooling terms.

While accounting  for the neutrino cooling term it is worth mentioning that the density and the 
temperature profiles of the disks that we are considering are not high enough so that the disk becomes
opaque to the neutrinos, i.e. we work in the `neutrino transparent limit'. Therefore in this situation 
the neutrino cooling rate is just the sum of four terms, 
\begin{equation}
Q_\nu ^- =( q_{eN} ^- + q_{e^+ e^-} ^- + q_{brem} ^- + q_{plasmon} ^- )H
\end{equation}
where $q_{eN} ^-$ is the cooling rate due to electron-positron pair capture onto free nucleons (``N" 
stands for protons and neutrons), $q_{e^+ e^-}$ is from electron positron pair 
annihilation, $ q_{brem} ^-$ is from nucleon-nucleon bremsstrahlung and $ q_{plasmon} 
^-$ is from plasmon decays. For the expressions of each of these cooling rates
see equations (47) to (54) of KNP05.

The forth term in equation (12) is the photodisintegration cooling rate which is given in equation (38), 
(39) and (40) of KNP05 and we do not repeat the relevant discussions here.

Last, $Q^-_{nuc}$ denotes the net energy released/absorbed by the nuclear reactions, other than 
photodisintegration. Note that for energy release it will be negative.

The density and temperature profiles for the accretion disks around a $3 M_\odot$ Schwarzschild black hole 
with $\dot{M}=0.01$ and $\dot{M}=0.001$ $M_\odot s^{-1}$are shown in Figure 1. In both
the cases, we have chosen the $\alpha$-viscosity to be 0.01. 

\section{Nucleosynthesis inside accretion disks}

\subsection{Initial composition of the infalling material}
In the collapsar II model, a mild supernova explosion is driven and a part of the supernova ejecta 
falls back onto the nascent neutron star, which 
then collapses into a black hole. In our model, we consider that the disk 
is made from the fallback material which may 
successfully drive a supernova. For accretion disks in Type II 
collapsars, the chemical composition of the fallback material undergoes a drastic change from the pre-SN 
composition due to SN shock heating. The chemical composition of the explosively burned O-rich layer is 
similar to the pre-SN composition of the Si-rich layer (Hashimoto 1995).
Similarly, material in the Si-rich layer changes its composition to that of the He-rich layer via explosive 
burning. Hence, for our present work, we take both Si-rich layer and He-rich layer at the outer boundary of 
the accretion disk. 

\begin{small}
\begin{LARGE}

\end{LARGE}
\end{small}

\begin{figure}
\centering
\includegraphics[width=1.1\columnwidth]{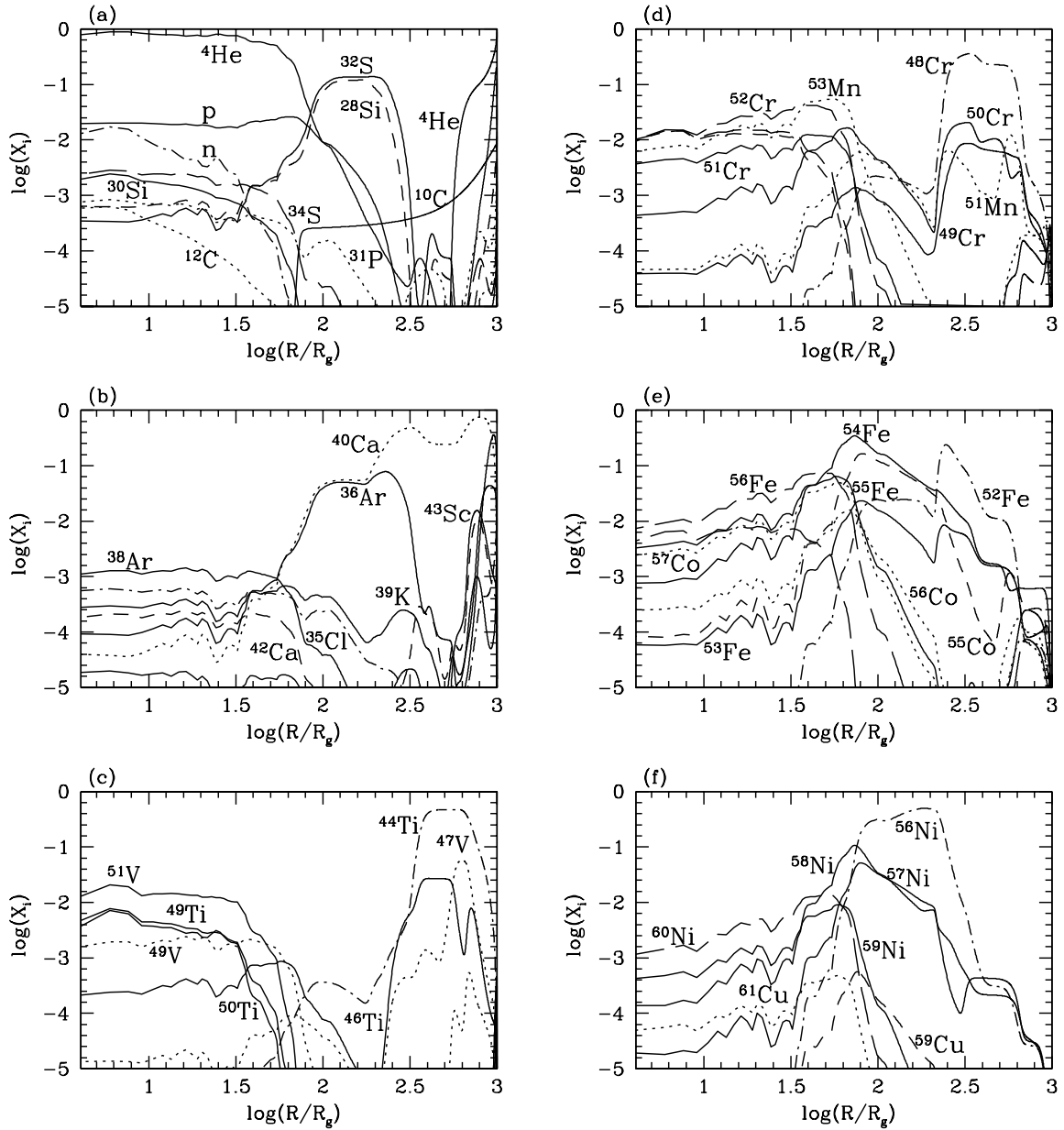}
\caption{
Abundance distribution inside the accretion disk for a $3 M_\odot$ Schwarzschild black hole with 
$\dot{M}=0.001M_\odot s^{-1}$, when pre-SN He-rich abundance is chosen at the outer disk.
 }
\label{figlam} \end{figure}

\begin{small}
\begin{LARGE}

\end{LARGE}
\end{small}

\begin{figure}
\centering
\includegraphics[width=0.9\columnwidth]{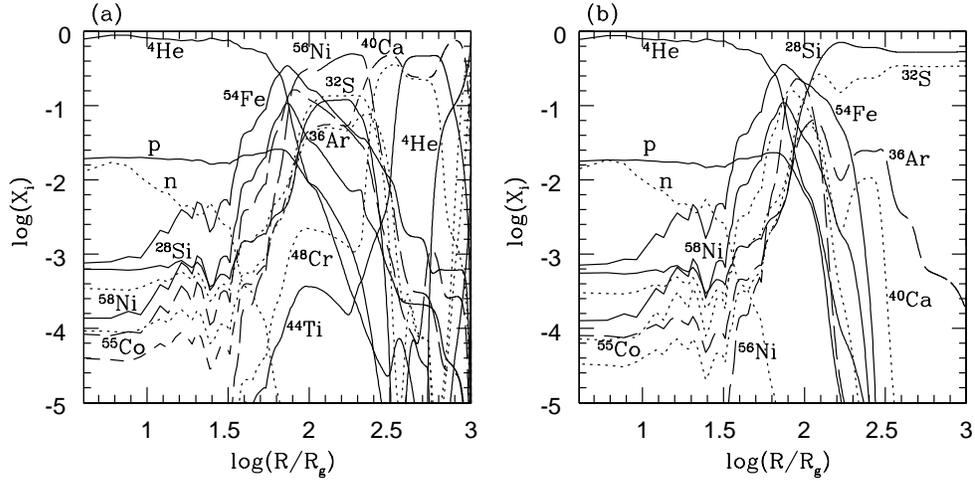}
\caption{Formation of zones characterized by dominant elements inside an  
accretion disk around a $3 M_\odot$ Schwarzschild black hole accreting at the rate of 
$\dot{M}=0.001M_\odot s^{-1}$ with (a) pre-SN He-rich abundance taken at the outer disk, (b) pre-SN 
Si-rich abundance taken at the outer disk.}
\label{figlam} \end{figure}
\subsection{Nuclear reaction network}

We use well tested nuclear network code, which has been implemented for more than two decades by 
Chakrabarti et al. (1987) and Mukhopadhyay \& Chakrabarti (2000) in the context of accretion disks. Cooper 
et al. (2006) used it to study superbursts on the neutron star surface. We have modified this code  
further by increasing the nuclear network and including reaction rates from the JINA Reaclib Database,
https://groups.nscl.msu.edu/jina/reaclib/db/.

\section{Abundance evolution inside accretion disks}
Once the disk model including the effects due to nuclear reactions is set, we calculate the density and 
temperature distribution inside the accretion disk for a given value of $\dot{M}, \alpha$ and $M$. 
Subsequently, we follow the abundance evolutions inside the disk as the material falls towards the black 
hole using the nuclear reaction network mentioned in \S 3.2.

\subsection{Model I}
This model consists of the situation when we take a Schwarzschild black hole of $M=3M_\odot$,  
$\dot{M}=0.001M_\odot s^{-1}$ and  $\alpha=0.01$. The composition of the accreting gas far from the black 
hole is set to be that of the explosively 
burned Si-rich layer which is similar to the pre-SN composition of the He-rich layer.

\subsubsection{Underlying reactions}

If we examine Fig. 2, we notice that many isotopes of several elements are synthesized in the
low accretion rate ($\dot{M}=0.001M_\odot s^{-1}$) He-rich disk. Elements which are synthesized more
than 1\%
in this disk are $^{28}$Si, $^{32}$S, $^{36}$Ar, $^{39}$K, $^{40}$Ca, $^{43}$Sc, $^{44}$Ti, $^{46}$Ti,
$^{47}$V,  $^{51}$V, $^{48}$Cr, $^{50}$Cr, $^{51}$Cr, $^{52}$Cr, $^{53}$Mn, $^{52}$Fe, $^{53}$Fe, 
$^{54}$Fe, $^{55}$Fe, $^{56}$Fe, $^{55}$Co, $^{56}$Co, $^{57}$Co, $^{56}$Ni, $^{57}$Ni, $^{58}$Ni, 
$^{59}$Ni, $^{60}$Ni. Apart from these, as is evident from Fig. 2, some amount of isotopes of copper are 
also
synthesized in this disk. Figure. 3(a) depicts the most abundant elements synthesized in the disk. If we
examine Fig. 3(a), we notice that the disk consists of several zones characterized by
the dominant elements. For example, the outermost region of the disk from 
$\sim 1000-800 R_{g}$ is mainly 
predominated by $^{40}$Ca. To begin with, the disk was rich in helium. It is important to remember that 
since the disk is made from the fallback material of the successfully driven supernova, its chemical 
composition is completely changed from the pre-SN composition. Therefore as mentioned in \S 3.1, the He-
rich 
layer is actually borne out of Si-rich layer via explosive burning (Hashimoto 1995). Thus, it is always 
possible that it may contain some unburnt $^{28}$Si, $^{32}$S and $^{36}$Ar. This $^{36}$Ar then undergoes
$\alpha$-capture reaction to give rise to $^{40}$Ca through $^{36}\rm Ar(\alpha,\gamma)^{40}Ca$. In fact, 
in
this region it is clear from Fig. 2(a) and Fig. 3(a) that there is substantial decline in the mass 
fraction of $^{4}$He.
Also in this zone $^{39}$K, $^{43}$Sc and $^{44}$Ti are synthesized more than 1\% which is evident from 
Fig. 2(b) and Fig. 2(c) respectively.

Next, between $\sim 800-600 R_{g}$, the mass fractions of $^{40}$Ca, $^{39}$K and $^{43}$Sc decline to some 
extent which is clear from Fig. 2(b) to give rise to $^{44}$Ti further which is shown in Fig. 2(c). 
$^{40}$Ca 
undergoes $\alpha$-capture 
reaction to give rise to $^{44}$Ti
via $^{40}\rm Ca(\alpha,\gamma)^{44}Ti$. Although $^{40}$Ca is the chief source of $^{44}$Ti in this part 
of 
the disk, $^{44}$Ti is also synthesized from $^{43}$Sc and $^{39}$K via $^{43}\rm Sc (p,\gamma)^{44}Ti$ and 
 $^{39}\rm K (p,\gamma)^{40}Ca(p,\gamma)^{41}Sc(\alpha,p)^{44}Ti$ respectively. Once the mass fraction of 
$^{44}$Ti becomes greater than the mass fraction of $^{40}$Ca, the mass fraction of $^{44}$Ti becomes
roughly constant for sometime, while the mass fraction of $^{40}$Ca keeps diminishing. 
In the mean time, the mass fraction of $^{48}$Cr starts rising. From Fig. 3(a),
it can be seen that this happens 
roughly between  $\sim 600-550 R_{g}$ and during this time $^{40}$Ca further undergoes a couple of 
$\alpha$-capture reactions: $^{40}\rm Ca (\alpha,\gamma)^{44}Ti(\alpha,\gamma)^{48}Cr$ resulting in 
$^{48}$Cr.
Once the mass fraction of $^{48}$Cr becomes nearly equal to the mass fraction of $^{40}$Ca, the
abundances of both the elements saturate. The abundance of $^{44}$Ti remains unchanged in this regime as
well. This continues for upto $\sim 425 R_{g}$ which is depicted in Fig. 3(a). 

Between $\sim 425-300 R_{g}$ (as shown in Figs. 2(b), 2(c) and 3(a)), the mass fraction of $^{44}$Ti 
declines 
while that of $^{40}$Ca gets enhanced.
This is due to the photodisintegration of $^{44}$Ti to $^{40}$Ca through the reaction: $^{44}\rm Ti(\gamma,
\alpha)^{40}Ca$. Therefore, roughly speaking the region between $\sim 1000-300 R_{g}$ is mainly the
$^{40}$Ca, $^{44}$Ti and $^{48}$Cr rich zone. 

Inside this region, the temperature and density in the disk are such that it favors complete 
photodisintegration 
of $^{44}$Ti and $^{48}$Cr releasing $\alpha$-particles and resulting in the formation of $^{40}$Ca,  
$^{36}$Ar, $^{32}$S and $^{28}$Si, as is evident from Fig. 2(a), Fig. 2(b) and Fig. 3(a). 
Subsequently, 
$^{28}$Si and $^{32}$S start burning, which favors formation of 
iron-group elements. This is because 
at high temperatures, as is in this case, $(\gamma,p)$, $(\gamma,\alpha)$ and $(\gamma,n)$ reactions and 
their
reverses come in equilibrium. But many of the photoejected particles are recaptured by nuclei where their
binding energy is high compared to the nuclei from where they are ejected. These reactions are often 
termed as {\it photodisintegration rearrangement} reactions in the literature (see, e.g., Clayton 1968) 
through 
which a 
rearrangement of loosely bound nucleons into more tightly bound states occur. 
As the rearrangement process proceeds, many elements are synthesized which are prone to $\beta-$decay, that 
in turn may increase the neutron to proton ratio of the nuclear gas.
The density and temperature regimes of the accretion disks under consideration are such that the 
photodisintegration
rearrangement time is much less compared to the $\beta-$decay time scale (Clayton 1968) and hence the total
neutron to proton ratio (free and bound) is very close to unity. Under such circumstances, the major 
iron group elements that can be synthesized are $^{54}$Fe or $^{56}$Ni, depending on the temperature of
the system. In our case, $^{56}$Ni is more dominant than $^{54}$Fe.
This is entirely attributed to the underlying hydrodynamics of the accretion disk (Clayton 1968). 
Apart from these {\it photodisintegration rearrangement} reactions, $(p,n), (\alpha,n), 
(\alpha,p)$ and 
their 
inverses also take place in these disks. The further details of the nuclear reactions in these disks
will be given in a subsequent work which will focus more on nuclear reactions (Banerjee \& Mukhopadhyay,
submitted).
Therefore, between $\sim 300-80 R_{g}$, as is evident from Fig. 3(a), there is a zone which is overabundant 
in $^{56}$Ni, 
$^{54}$Fe, $^{32}$S and 
$^{28}$Si. Inside this zone, all the heavy elements photodisintegrate to $^{4}$He, neutron and proton.
Had the temperature in the disk been higher than what is achieved here, (which would be the case of
$\dot{M} \sim 0.01 M_{\odot} s^{-1}$), then even $^{4}$He would have been 
photodisintegrated to protons and neutrons.
Thus, from $\sim 80 R_{g}$ to the inner edge of the disk, there is the $^{4}$He rich 
zone (see, e.g., Fig. 2(a) and Fig. 3(a)). 

\subsection{Model II}
\begin{small}
\begin{LARGE}

\end{LARGE}
\end{small}

\begin{figure}
\centering
\includegraphics[width=1.1\columnwidth]{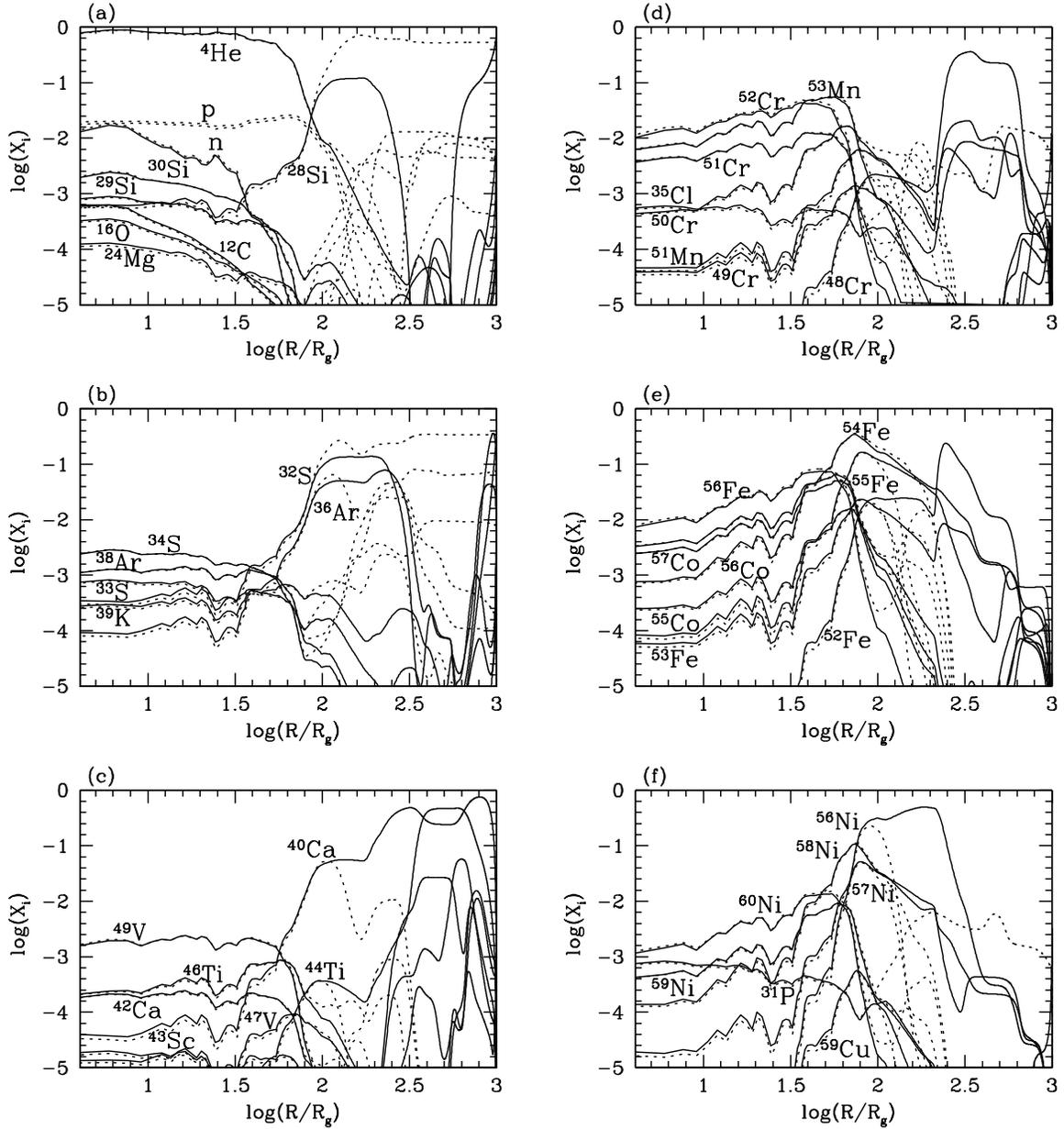}
\caption
{Comparison of the abundance of various elements between He-rich and Si-rich disks with $\dot{M}=0.001 
M_\odot s^{-1}$ and $M= 3 M_{\odot}$,
where the solid lines 
represent the abundance of elements of the He-rich disk and the dotted lines represent that of the Si-rich 
disk.}
\label{figlam} \end{figure}

In this model, we take a $3M_\odot$ Schwarzschild black hole with accretion rate $\dot{M}=0.001 M_\odot 
s^{-1}$ and $\alpha$ viscosity being $0.01$ as before. However, the composition of the accreting gas far 
from the black hole is set to be that of the explosively-burned O-rich layer which is similar to the pre-SN 
composition of the Si-rich layer. 

\subsubsection{Underlying reactions}

This disk has some similarities and some dissimilarities in the abundance evolution pattern compared to the 
above mentioned He-rich disk. The differences lie mainly in the outer disk, outside of, say, $R\sim 100 
R_{g}$.
This is quite clear from Fig. 4. If we examine Fig. 3(b), we notice that it has a huge
zone rich in $^{28}$Si and $^{32}$S at the outer disk. This is because we begin with a $^{28}$Si and 
$^{32}$S rich outer disk and the disk hydrodynamics does not favor silicon burning upto $R \sim 250 R_{g}$.
Inside this radius, silicon burning commences and soon the disk becomes rich in $^{54}$Fe, $^{56}$Ni and
$^{58}$Ni. Thus there is a very narrow zone rich in these elements in this disk which is clear from
Fig. 3(b). Inside $\sim 70 R_{g}$, all the
heavy elements again get photodisintegrated to $^{4}$He, protons and neutrons. Thus, from  $\sim 70 R_{g}$ 
inwards, there is the He-rich zone (see, e.g., Fig 3(b)). 

Another very remarkable feature in the He-rich and Si-rich disks is that inside  $\sim 100 R_{g}$, the 
abundance of various elements start becoming almost identical as if once a threshold density and 
temperature is achieved, the nuclear reaction rates forget their initial abundance and follow only the 
underlying disk hydrodynamics. One possible explanation for this is that, once Si-burning starts taking 
place in these disks the underlying disks in both the 
cases
have the same hydrodynamics, and hence the final end products
are always iron-group elements. Finally in the inner disk, it is obvious that all the heavy elements will
photodisintegrate to $\alpha$ particles and if $\dot{M}$ is higher than what is considered in 
these 
cases, the $\alpha$ particles will further breakdown into protons and neutrons. 

\subsection{High accretion cases}
If we increase $\dot{M}$ ten times around the same $3M_\odot$ Schwarzschild black hole and 
again consider He-rich and Si-rich abundances at the outer disk we find that the individual zones
shift outward retaining similar composition as is in the low $\dot{M}$ cases described above. The outermost 
zone in
these cases is the
zone rich in $^{54}$Fe and $^{56}$Ni. Another important feature in these disks is that for  
$R \lsim 500 R_{g}$
the He-rich and the Si-rich disks start looking identical in terms of elemental abundances.

\section{Summary}
We have studied the nucleosynthesis inside accretion disks associated with the fallback collapsars using 
1.5-dimensional disk model and the nuclear reaction network. The syntheses of isotopes of iron, cobalt,
nickel, silicon, sulphur and argon 
have been reported in the disk, as was reported previously by others. However, apart from these elements, 
we have shown, for the first time in the literature to the best of our knowledge, that several unusual 
nuclei like $^{31}$P, $^{39}$K, $^{43}$Sc, $^{35}$Cl, 
and various uncommon
isotopes of titanium, vanadium, chromium, manganese and copper get synthesized in the disk. 
We have discussed that several zones characterized by dominant elements are formed in the disk. In fact 
once a threshold density and temperature is reached the elemental distribution in the disk
starts looking identical. The respective zones and the radius from where the elemental abundance starts 
looking identical shift outwards on increasing the mass accretion rate.  

Huge fractions of 
these elements thus synthesized are expected to be ejected from the disk via outflows (e.g. Surman et al. 2006).
Unfortunately, while emission lines of many of these 
elements were discovered in the X-ray afterglows of GRBs by BeppoSAX, Chandra, XMM-Newton etc., 
e.g. Fe lines in GRB 970508 (Piro et al. 1999), 
GRB 970828 (Yoshida 
et al. 1999) and GRB 000214 (Antonelli et al. 2000); S, Ar, Ca lines in GRB 011211 (Reeves et al. 2003),
Swift appears not to have found these lines (e.g. Zhang et al. 2006; Hurkett et al. 2008) yet.
We however synthesize all these elements in our accretion disks. 

In our upcoming work, we plan to consider nucleosynthesis in the outflow from these disks in detail 
(Banerjee \& Mukhopadhyay, submitted; also see Banerjee \& Mukhopadhyay 2013). 
Note that for the Type~II collapsar model considered in the present work, magnetohydrodynamic 
processes are more plausible mechanism for outflows and jets (see, e.g., MacFadyen et al. 2001).
It was found that for the accretion rate $\sim 0.001-0.01 M_\odot$s$^{-1}$, the available 
energy is $\sim 10^{51}-10^{52}$ ergs, which potentially could produce jets.
This conservative energy estimate by previous authors (MacFadyen et al. 2001)
is still large compared to the energy of a typical supernova. Therefore, the matter 
will have enough energy to make a powerful GRB, if it is collimated
into a small fraction of the sky.
Depending on which of the elements survive in the outflow, we can then 
predict how far they match with observations and whether the disk outflow will lead to a supernova 
explosion or not. 

\section*{Acknowledgments}

\indent \indent The authors would like to thank the referee for useful comments, particularly in the 
context of Swift data. This work was partly supported by the ISRO grant ISRO/RES/2/367/10-11.

\end{document}